\documentclass[a4paper,pre,twocolumn,superscriptaddress,showpacs]{revtex4-1}

\usepackage[utf8]{inputenc}
\usepackage[dvips]{graphicx}
\usepackage{natbib}
\usepackage{amssymb}
\usepackage{latexsym}
\usepackage{amsmath}

\usepackage{longtable} 
\usepackage{rotating}
\usepackage{multirow}

\newcommand{\eA}{\epsilon_A}

\begin{document}

%\begin{frontmatter}

\bibliographystyle{elsarticle-harv}

\title{Evolutionary stability and resistance to cheating in an indirect
reciprocity model based on reputation}

\author{Luis A. Martinez-Vaquero}
\affiliation{Grupo Interdisciplinar de Sistemas Complejos (GISC),
Departamento de Matem\'aticas, Universidad Carlos III de Madrid,
28911 Legan\'es, Madrid, Spain}
\author{Jos\'e A. Cuesta}
\affiliation{Grupo Interdisciplinar de Sistemas Complejos (GISC),
Departamento de Matem\'aticas, Universidad Carlos III de Madrid,
28911 Legan\'es, Madrid, Spain}
\affiliation{Instituto de Biocomputaci\'on y F\'\i sica de Sistemas
Complejos (BIFI), Universidad de Zaragoza, 50009 Zaragoza, Spain}

%\author[1]{Luis A.\ Martinez-Vaquero}
%\ead{luisalberto.martinez@uc3m.es}
%\author[1,2]{Jos\'e A.\ Cuesta\corref{cor1}}
%\ead{cuesta@math.uc3m.es}

%\address[1]{Grupo Interdisciplinar de Sistemas Complejos (GISC),
%Departamento de Matem\'aticas, Universidad Carlos III de Madrid,
%Avenida de la Universidad 30, 28911 Legan\'es, Madrid, Spain}

%\address[2]{Instituto de Biocomputaci\'on y F\'\i sica de Sistemas Complejos (BIFI),
%Universidad de Zaragoza, Campus R\'\i o Ebro, 50018 Zaragoza, Spain}


%\date{\today}

%\begin{document}

\begin{abstract}
Indirect reciprocity is one of the main mechanisms to explain the emergence and
sustainment of altruism in societies. The standard approach to indirect
reciprocity are reputation models. These are games in which players base their
decisions on their opponent's reputation gained in past interactions with other
players (moral assessment). The combination of actions and moral assessment
leads to a large diversity of strategies, thus determining the stability of any
of them against invasions by all the others is a difficult task. We use a
variant of a previously introduced reputation-based model that let us
systematically analyze all these invasions and determine which ones are
successful. Accordingly we are able to identify the third-order strategies
(those which, apart from the action, judge considering both the reputation
of the donor and that of the recipient) that are evolutionarily stable. Our
results reveal that if a strategy resists the invasion of any other one sharing
its same moral assessment, it can resist the invasion of any other strategy.
However, if actions are not always witnessed, cheaters (i.e., individuals with
a probability of defecting regardless of the opponent's reputation) have a
chance to defeat the stable strategies for some choices of the probabilities of
cheating and of being witnessed. Remarkably, by analyzing this issue with
adaptive dynamics we find that whether a honest population resists the invasion
of cheaters is determined by a Hamilton-like rule---with the probability that
the cheat is discovered playing the role of the relatedness parameter.
\end{abstract}

\pacs{02.50.-r,87.10.-e,87.23.-n,89.75.Fb}

\maketitle

%\begin{keyword}

%\end{keyword}

%\end{frontmatter}

\section{Introduction}

Human being is the social animal par excellence. An individual can help another
even if it is the first time they meet or if they know that they will never meet
again. Several mechanisms have been proposed to explain cooperation between
unrelated individuals. Among them reciprocity, either direct or indirect, stands
as one of the most successful explanations of altruism \cite{fehr:2003}. In direct
reciprocity individuals pay back the help received in repeated encounters with
the same partner (``I help you if you help me'') \cite{trivers:1971}. In
society, however, many interactions have low chances to be repeated with the
same individual. To explain altruism in those interactions,
the concept of indirect reciprocity was introduced
\cite{sugden:1986, alexander:1987}.
Through this mechanism, individuals do not receive the consequences of their
actions directly from the individuals they interact with, but indirectly
through society (``I help others to be helped by others''). Indirect
reciprocity is an important mechanism for the emergence and sustainment of
altruism not only in small-scale human societies \cite{dufwenberg:2001,
milinski:2002, panchanathan:2004, semmann:2004, suzuki:2007} but in other
species as well \cite{bshary:2006}. And it certainly plays an important role
in communication networks \cite{bolton:2004, keser:2002}.

There are two types of indirect reciprocity: upstream and downstream. In
upstream reciprocity \cite{boyd:1989, nowak:2007} an individual opts for a
given action taking into account if she was previously helped or not. In
this respect upstream reciprocity is more akin to a learning mechanism,
because individuals adapt their choices based on their past experience.
In downstream reciprocity---also called reputation-based indirect
reciprocity---an individual assigns a reputation to the others taking into
account how they interact with the rest of the society \cite{nowak:1998,
ohtsuki:2004, brandt:2004, nowak:2005, milinski:2002}. These reputations allow
her to decide whether she should help these individuals or not in potential
future encounters with them. Accordingly, downstream indirect reciprocity
is a cognitively very demanding task: it requires observation, memory and
communication. It is this reputation-based indirect reciprocity that will be the
focus of the present work.

Two different kinds of models of reputation-based indirect reciprocity have
been considered in the literature. In indirect observation models
\cite{ohtsuki:2004} each action is observed and judged only by one individual,
who spreads this information across the population through verbal communication
and gossip. Therefore all individuals share the same opinion about each other.
On the contrary, in direct observation models \cite{leimar:2001, brandt:2004,
pacheco:2006} everyone witnesses the action and makes her private judgment of
it. Thus individuals' different opinions about the rest of the members of the
society can coexist in this kind of models.

Ohtsuki and Iwasa \citep{ohtsuki:2004} and Brandt and Sigmund \cite{brandt:2004}
have proposed a classification of the different strategies in games with
indirect reciprocity through their assessment and action modules. Strategies
can be classified either as second order or as third order strategies. In both
cases, the reputation is assigned taking into account the observed action and
the reputation of the individual who received it. But third-order strategies
also look at the reputation of the individual who performs the action. The
dynamics of second-order assessments has been explored in \cite{ohtsuki:2007b}.
Ohtsuki and Iwasa \citep{ohtsuki:2004} also studied systematically the
evolutionarily stability of third order strategies. Their model is an indirect
observation model and therefore the whole society shares the same moral
assessment.  Stability is studied by confronting strategies with different
action rules.  They concluded that there are eight strategies---the so-called
\textit{leading eight}---which are evolutionary stable strategies (ESS) under
these assumptions. The meaning and success of these strategies has also been
studied by Ohtsuki and Iwasa \cite{ohtsuki:2006b}. On the other hand, Uchida
and Sigmund \cite{uchida:2010} have chosen some of the leading eight strategies
that share the same action rules but have different moral assessment and have
confronted them in a model with private opinions.

In this work, we extend the systematic study carried out by Ohtsuki and Iwasa
confronting strategies with different moral assessments. Unlike their work,
we use a direct observation model in which individuals no longer share the same
opinion about the rest of the population. We introduce the concept of coherence
as a measure of the relation between the moral assessment and the action rules
of a strategy and study how it relates to the stability and efficiency of the
strategies. We identify which strategies resist the invasion of all the other
strategies, i.e., which combinations of moral assessment and action rules
emerge under this evolutionary competition. Finally we explore the effect that
an action is witnessed by nobody in the population. Individuals can then face
the risk to cheat---i.e., defect regardless of the opponent's reputation---at
no own reputation cost.

The present paper is structured as follows. In section~\ref{sec:model} we
introduce the model. In section~\ref{sec:implementation} we describe its
mathematical implementation. We study homogeneous populations and discuss their
stability against invasions by other strategies. We also analyze the effect on
introducing a probability of cheating, when actions have a chance not to be
witnessed. Finally, our results are shown in section~\ref{sec:results} and
discussed in section~\ref{sec:discussion}.

\section{Model}
\label{sec:model}

Brandt and Sigmund \cite{brandt:2004} introduced a very stylized model of
indirect reciprocity based on reputation, and Ohtsuki and Iwasa
\citep{ohtsuki:2004, ohtsuki:2006b} investigated the stability of its
strategies under the assumption that all individuals share the same moral
judgment.

The model we will be dealing with in this work is a slight modification of this
basic model. It consists of an infinite, well-mixed population, of interacting
and judging individuals. Every time step a pair of individuals are randomly and
equiprobably drawn from the population. One of them plays the role of the
\emph{donor} and the other one of the \emph{recipient}. The donor then decides
whether to pay a cost $c>0$ to help (C) the recipient or not (D). If the
recipient is helped, she receives a benefit $b>c$. This \emph{action} is
observed by every individual of the population (including themselves).
Observers privately judge the donor for the action taken on the recipient
according to their own \emph{moral assessment,} and assign her a
reputation---either good (G) or bad (B)---accordingly. Therefore every
individual in the population has a private opinion of every other individual,
including herself.

This process is repeated until the population reaches an equilibrium (we will
define this equilibrium in more precise terms in the next section). Then the
average payoff that every individual receives in this repeated game is
computed. Direct reciprocity is excluded from this game because the population
is virtually infinite---hence the probability that two people meet again is
negligible.

We consider third order indirect reciprocity, i.e., each strategy is described
by two moduli: the action rules and the moral assessments.

The action rules determine what the donor must do (either help or refuse to
help) given the reputation of both players. Specifically, $a_{i\alpha\beta}=1$
(C) if strategist $i$ with reputation $\alpha$ helps an individual with
reputation $\beta$ (both according to $i$'s moral judgments) and $0$ (D)
otherwise.

The moral assessments tell the individual if the action just
witnessed should be judged as good or bad, hence revising the donor's
reputation. Specifically, $m_{i\alpha\beta}(a)=1$ (G) if strategist $i$ assigns
good reputation to a donor previously judged $\alpha$ by $i$, who performs an
action $a$ on an recipient previously judged $\beta$ by $i$, and is $0$ (B)
otherwise.

Thus each strategy is defined by 12 numbers: 4 for the action module and 8 for
the moral module. This amounts to 4096 different possible strategies. Although
a thorough study of mutual invasions and coexistence of different strategies
---as that performed in Ref.~\cite{martinez:2012} for a direct reciprocity
model--- would be desirable, the wealth of strategies forbids it, and we
should content ourselves with a pairwise test of mutual invadability.

We will assume that sometimes players do not act according to their action
rules \cite{leimar:2001, ohtsuki:2004, panchanathan:2003, fishman:2003,
lotem:1999}. Thus, with a probability $\eA$ a donor defects regardless of her
action rules and with $1-\eA$ she performs the action she planned to. Another
source of errors is misjudgment, i.e., and individual can make a mistake in
interpreting the action. In this category lies social pressure.  This is a kind
of error that is especially important if the information on the action
performed is spread by gossiping, because then, a misjudgment of the witness
will lead to a misjudgment of the entire population. Otherwise, it affects only
a small fraction of the individuals. Since keeping track of errors may lead to
a proliferation of judgments---even between individuals sharing the same moral
assessment---and render the model computationally unfeasible, we will content
ourselves by implementing only errors in the action.

\section{Mathematical implementation of the model}
\label{sec:implementation}

\subsection{Homogeneous populations}

Let us start by assuming that there is only one strategy $i$ present in the
population. Let $x_{i}$ be the fraction of individuals considered good by the
whole population (there is a unique moral assessment). Then the rate of change
of $x_{i}$ is given by 
\begin{equation}
\frac{dx_{i}}{dt}=\sum_{\alpha\beta} \chi_\alpha(x_{i}) \chi_\beta(x_{i})
P_{i,\alpha\beta}-x_{i},
\label{eq:homo}
\end{equation}
where $P_{i,\alpha\beta}$ is the probability that a donor of reputation
$\alpha$ acting on a recipient with reputation $\beta$ is considered good
by the population. This probability can be obtained as
\begin{equation}
P_{i,\alpha\beta} =(1-\eA)m_{i\alpha\beta}(a_{i\alpha\beta})+\eA
m_{i\alpha\beta}(D)
\label{eq:P-homo}
\end{equation}
because with probability $\eA$ no help is provided and with probability $1-\eA$
the action performed is $a_{i\alpha\beta}$, as prescribed by the action module.
We have also introduced the auxiliary function $\chi_{\gamma}(x_i)$,
\begin{equation}
\chi_{\gamma}(x_i)=\gamma x_i+(1-\gamma)(1-x_i),
\end{equation}
which in this case represents the fraction of individuals with reputation
$\gamma$.

The dynamics reaches an equilibrium when $x_i=\sum_{\alpha\beta} \chi_\alpha(x_i)
\chi_\beta(x_i) P_{i,\alpha\beta}$. Therefore the fraction of good individuals in a
homogeneous population in equilibrium is the solution $0\leqslant x_i\leqslant 1$
of the quadratic equation $F(x_i)=0$, where
\begin{equation}
\begin{split}
F(x_i) =& x_i^2(P_{i,11}+P_{i,00}-P_{i,10}-P_{i,01}) \\
     &+x_i(P_{i,10}+P_{i,01}-2P_{i,00}-1) +P_{i,00}.
\end{split}
\label{eq:2grade}
\end{equation}
As $F(0)=P_{i,00}\geqslant0$ and  $F(1)=P_{i,11}-1\leqslant 0$, there is always a
solution in $[0,1]$, but in some cases there may be two (when $P_{i,00}=0$ or
$P_{i,11}=1$ or both), one stable and one unstable, and there is a degenerate
case (when all coefficients in $F(x_i)$ vanish) in which any $x_i$ is a solution.
In this latter case, adding a small error, $\epsilon_m$, in the moral
assessment determines uniquely a stable solution. When the population is
homogeneous this can be done at no computational cost by simply replacing
$P_{i,\alpha\beta}$ in Eq.~\eqref{eq:2grade} by
$(1-2\epsilon_m)P_{i,\alpha\beta}+\epsilon_m$. This yields the expression
$F(x_i)=\epsilon_m(1-2x_i)$, whose only root is $x_i=1/2$, regardless of
$\epsilon_m$. Hence we take this solution---which holds even in the limit
$\epsilon_m\to 0$---as the solution of this degenerate case.

Given the equilibrium fraction $x_{iH}$, the probability that an individual helps
another is
\begin{equation}
\theta_{iH}=(1-\epsilon_a)
\sum_{\alpha\beta} \chi_\alpha(x_{iH})\chi_\beta(x_{iH})a_{i\alpha\beta}.
\end{equation}
Therefore the average payoff that any individual in this population obtains is
\begin{equation}
W_{iH}=(b-c)\theta_{iH}.
\end{equation}
As the whole population shares the same strategy, it can be regarded a measure
of `self-efficiency'. This provides a mean to classify strategies.

\emph{Coherence} provides an alternative classification criterion. Given an
action $a$ that a donor with reputation $\alpha$ performs on a recipient
with reputation $\beta$, we call an individual coherent if placed on the
donor's feet she performs the same action $a$ when she morally assesses it
as good, and the opposite action $1-a$ when she morally assesses it as bad. In
other words, an individual is coherent if she performs actions that she judges
as good and do the opposite of actions that she judges as bad. Thus we can
introduce a coherency index $h$ as
\begin{equation}
 h_i=\frac{1}{2} \sum_{\alpha\beta\,a} \big[ 1- \left|
m_{i\alpha\beta}(a) - \delta(a,a_{i\alpha\beta}) \right| \big]
\chi_\alpha\left(x_{iH}\right) \chi_\beta\left(x_{iH}\right),
\label{eq:coherency}
\end{equation}
where $\delta(x,y)=1$ if $x=y$ and $0$ otherwise.
This index can range from 0 (no coherence) to 1 (full coherence). Notice that
the coherence of a strategy can change when more strategies are present in the
population, because it depends on the fraction of good and bad individuals.
Nevertheless, for the sake of classification, we have defined this index for a
homogeneous population so that it is uniquely determined by $x_{iH}$, and
therefore is an intrinsic feature of each strategy.
 
\subsection{Stability of strategies}
\label{sec:stability}

Consider now a homogeneous population where individuals share the same
\textit{resident} strategy. From time to time a small fraction of the
population can adopt a new \textit{mutant} strategy. This mutant strategy will
eventually invade the resident population if mutants obtain a higher payoff
than residents.

Calculating these payoffs requires to compute the four fractions of individuals
that are considered good and bad by the first and the second strategy in
equilibrium. In the limit where the fraction of mutants is very small both
residents and mutants interact only with residents. The dynamics of these four
fractions of individuals is given in this limit by the equations
\begin{align}
\frac{dx_{1}^{\Lambda_1\Lambda_2}}{dt} =&
\sum_{\substack{\alpha_1\alpha_2 \\ \beta_1\beta_2}}
x_1^{\alpha_1\alpha_2}x_1^{\beta_1\beta_2}
P^{\Lambda_1\Lambda_2}_{1,\alpha_1\beta_1,\alpha_2\beta_2} -
x_{1}^{\Lambda_1\Lambda_2},
\label{eq:evol_x1} \\
\frac{dx_{2}^{\Lambda_1\Lambda_2}}{dt} =&
\sum_{\substack{\alpha_1\alpha_2 \\ \beta_1\beta_2}}
x_2^{\alpha_1\alpha_2}x_1^{\beta_1\beta_2}
P^{\Lambda_1\Lambda_2}_{2,\alpha_1\beta_1,\alpha_2\beta_2} -
x_{2}^{\Lambda_1\Lambda_2},
\label{eq:evol_x2}
\end{align}
where $x_{i}^{\Lambda_1\Lambda_2}$ are the fractions of $i$-strategists ($i=1$
for residents and $i=2$ for mutants) who are judged $\Lambda_1$ by residents
and $\Lambda_2$ by mutants; $P^{\Lambda_1\Lambda_2}_{i,\alpha_1\beta_1,
\alpha_2\beta_2}$ is the probability that an $i$-strategist with reputation
$\alpha_1$ for residents and $\alpha_2$ for mutants, acting on a recipient of
the resident population with reputation $\beta_1$ for other residents and
$\beta_2$ for mutants, is judged $\Lambda_1$ by residents and $\Lambda_2$ by
mutants. The form of this probability is
\begin{equation}
\begin{split}
P^{\Lambda_1\Lambda_2}_{i,\alpha_1\beta_1,\alpha_2\beta_2} = & (1-\eA)
\delta\big(\Lambda_1,m_{1\alpha_1\beta_1}(a_{i\alpha_i\beta_i})\big)\\
&\times
\delta\big(\Lambda_2,m_{2\alpha_2\beta_2}(a_{i\alpha_i\beta_i})\big) \\
 & +\eA \delta\big(\Lambda_1,m_{1\alpha_1\beta_1}(D)\big)\\
&\times
        \delta\big(\Lambda_2,m_{2\alpha_2\beta_2}(D)\big).
\end{split}
\end{equation}

Equations \eqref{eq:evol_x1} and \eqref{eq:evol_x2} can be simplified in the
equilibrium. Nonetheless some of the equations need to be numerically
solved (see Appendix A). To this purpose we must start from a sensible initial
condition. We will assume that just before the invasion begins, all
individuals---both mutants and resident---share the same opinion about
everybody. The rationale for this choice is that, before the change of strategy
undergone by mutants takes place, the population was homogeneous. Therefore
$x_i^{GG}(0)=x_{iH}$, $x_i^{BB}(0)=1-x_{iH}$ and $x_i^{GB}(0)=x_i^{BG}(0)=0$.

Once the fractions in equilibrium $x_{i}^{\Lambda_1\Lambda_2}$ are known, the
probabilities $\theta_{i,j}$ that an $i$-strategist helps a $j$-strategist
($i,j=1,2$) are obtained as
\begin{equation}
\begin{split}
\theta_{1,j} &=(1-\eA) \sum_{\alpha\beta} \chi_\alpha(x_{1}^{G*})\,
\chi_\beta(x_{j}^{G*})\, a_{1\alpha\beta}, \\
\theta_{2,j} &=(1-\eA) \sum_{\alpha\beta} \chi_\alpha(x_{2}^{*G})\,
\chi_\beta(x_{j}^{*G})\, a_{2\alpha\beta},
\end{split}
\label{eq_theta}
\end{equation}
where we have introduced the short-hand notation $x_i^{G*}=\sum_{\Lambda_2}
x_i^{G\Lambda_2}$ and $x_i^{*G}=\sum_{\Lambda_1} x_i^{\Lambda_1 G}$ to denote
the sum over a given reputation. Obviously, $x_i^{G*}$ ($x_i^{*G}$) is the
fraction of $i$-strategists that are judged as good by the resident (mutant)
players irrespective of the mutant's (resident's) judgement.

Finally, the average payoff $W(i|j)$ that an $i$-strategist receives from a
$j$-strategist can be computed as
\begin{equation}
W(i|j)=
\begin{cases}
(b-c)\, \theta_{i,i}, & i=j, \\
b\,\theta_{j,i} -c\,\theta_{i,j}, & i\ne j.
\end{cases}
\label{eq_W22}
\end{equation}

The resident population cannot be invaded by the mutants if $W(1|1)>W(2|1)$ or
if $W(1|1)=W(2|1)$ and $W(1|2)>W(2|2)$. If the resident strategy resists the
invasions of all the other mutant strategies it is considered evolutionarily
stable.

\section{Results}
\label{sec:results}

\subsection{Stability of strategies}

\begin{figure*}[!ht]
\begin{center}
\includegraphics[width=7in]{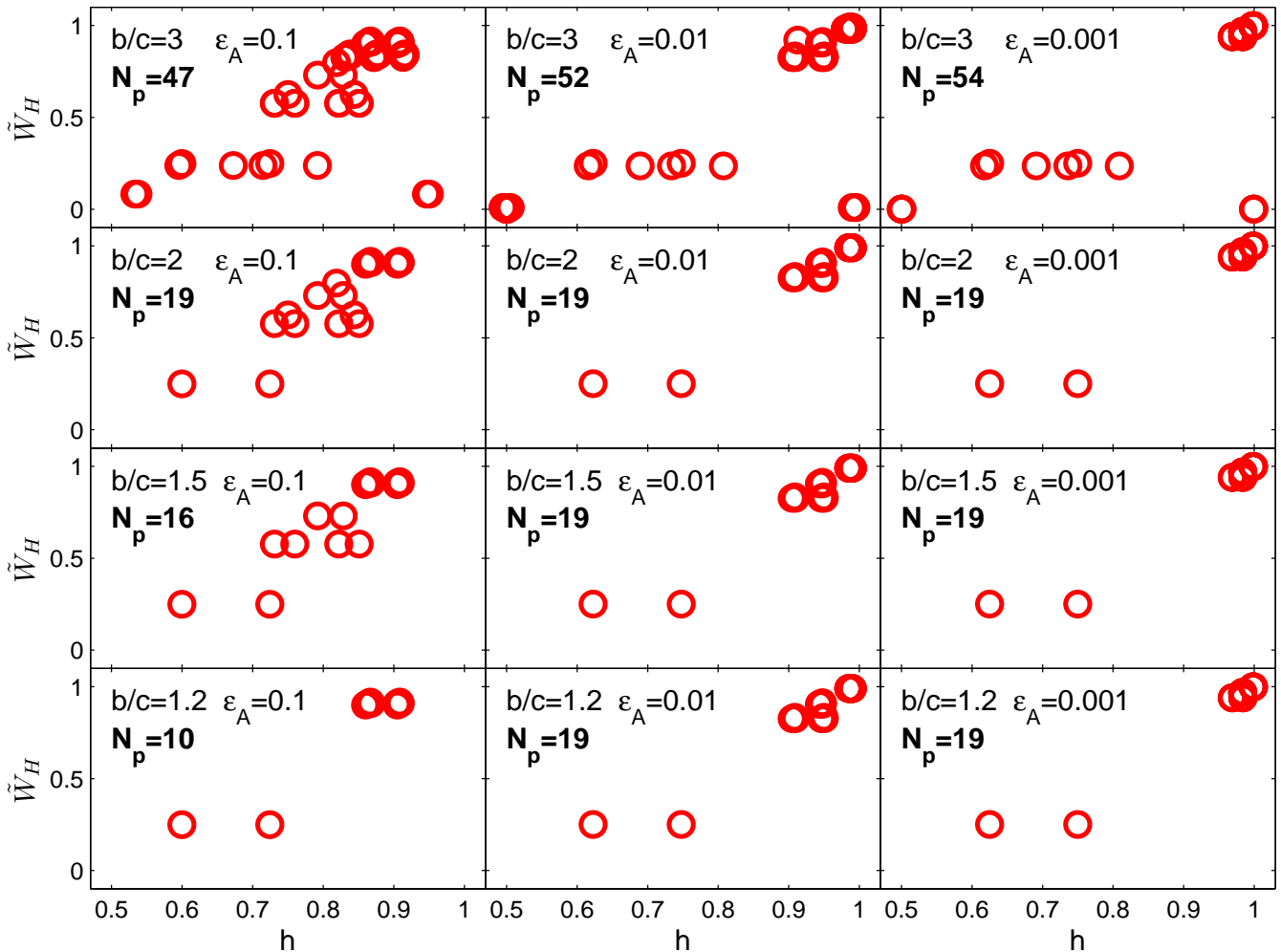}
\caption{(Color online) Representation of the normalized average payoff
$\widetilde{W}_H$ as a function of the coherence $h$ for the stables
strategies. Only the coherent
strategy ($h>0.5$) of each pair is represented. Different panels show results
for different $\eA$ and $b/c$.
\label{fig:samemoral}}
\end{center}
\end{figure*}

Our aim is to identify strategies that are evolutionarily stable. In principle
this requires for every strategy to check whether it can be invaded by every
other strategy. However the number of pairs of strategies is larger than
$1.5\times 10^7$, so this becomes too demanding a computational task.
Accordingly we proceed in two steps: (i) we look for all strategies that are
stable against invasions by other strategies sharing the same moral assessment;
and (ii) we study the stability of these selected strategies against all the
remaining ones.

Our Eqs.~\eqref{eq:evol_x1} and \eqref{eq:evol_x2} reduce to those used in
Ref.~\cite{ohtsuki:2004} if we fix the moral assessments and neglect moral
errors. We carried out our analysis for different values of the action error
$\eA$ ($0.1$, $0.01$ and $0.001$) and benefit-to-cost $b/c$ ratio ($1.2$,
$1.5$, $2$ and $3$).

In Fig.~\ref{fig:samemoral} we represent the strategies that are stable
against invasions by all strategies sharing the same moral assessment, as a
function of their normalized average payoff
$\widetilde{W}_H=W_H\,[(b-c)(1-\eA)]^{-1}$ and their coherence. These
strategies always appear in pairs since there is a symmetry in the reputation:
if labels ``good'' and ``bad'' are exchanged the results are not affected
(see~\cite{ohtsuki:2004} for more details).  Notice though that there is
symmetry only in the moral assessment but not in the action. The reason is that
cooperating and defecting are not just labels because they have consequences in
the payoffs obtained. It is easy to show, using Eq.~\eqref{eq:coherency} that
the sum of the coherences of a strategy and its ``mirror'' strategy is always
$1$. Coherence thus provides an external assessment on moral labels, breaking
the symmetry and permitting to differentiate between a strategy and its mirror.
In Fig.~\ref{fig:samemoral} we only show the results for the coherent
strategy ($h\geqslant 0.5$) of the pair and report how many pairs $N_p$ are
shown.

Figure~\ref{fig:samemoral} shows that the larger the benefit-to-cost ratio, the
higher the number of stable strategies; in other words, it is difficult to break
into a population whose individuals obtain high rewards for help. Moreover, we
have counted the number of pairs of strategies in which each strategy can be
invaded by the other---i.e., at least one mixed equilibrium is formed. The
number of these pairs also appears to be larger the higher the benefit-to-cost
ratio ($2500$--$2600$ pairs for $b/c=3$ vs.~$1500$--$1700$ for the remaining
cases). Therefore even if a mutant invades a resident strategy, it is less
likely that it eventually dominates the population if $b/c$ is high. From
Fig.~\ref{fig:samemoral} we also conclude that a high $\eA$ allows
invaders to spread easier in the resident population.

\begin{table*}[!ht] 
  \begin{center}
    \begin{tabular}{|ccccccccccccc|c|}
     \hline
	& $m_{GG}(C)$ & $m_{GG}(D)$ & $m_{GB}(C)$ & $m_{GB}(D)$ & $m_{BG}(C)$ &
$m_{BG}(D)$ & $m_{BB}(C)$ & $m_{BB}(D)$ & $a_{GG}$ & $a_{GB}$ & $a_{BG}$ &
$a_{BB}$ & $\widetilde{W}_H$ \\
     \hline
  Ia  & G & B & G & G & G & B & G & B & C & D & C & C & 0.9902 \\
  Ib  & G & B & B & G & G & B & G & B & C & D & C & C & 0.9902  \\
  IIa & G & B & G & G & G & B & G & G & C & D & C & D & 0.9901 \\
  IIb & G & B & G & G & G & B & B & G & C & D & C & D & 0.9901 \\
  IIc & G & B & B & G & G & B & G & G & C & D & C & D & 0.9901 \\
  IId & G & B & B & G & G & B & B & G & C & D & C & D & 0.9901 \\
  IIIa& G & B & G & G & G & B & B & B & C & D & C & D & 0.9900 \\
  IIIb& G & B & B & G & G & B & B & B & C & D & C & D & 0.9900 \\
      \hline
      & G & B & B & B & G & B & G & B & C & D & C & C & 0.9135 \\
      & G & B & B & B & G & B & G & G & C & D & C & D & 0.9049 \\
      & G & B & B & B & G & B & B & G & C & D & C & D & 0.9049 \\
      & G & B & B & G & B & B & G & B & C & D & D & C & 0.8340 \\
      & G & B & G & G & B & B & G & B & C & D & D & C & 0.8340 \\
      & G & B & B & G & B & B & B & G & C & D & D & D & 0.8264 \\
      & G & B & B & G & B & B & G & G & C & D & D & D & 0.8264 \\
      & G & B & G & G & B & B & B & G & C & D & D & D & 0.8264 \\
      & G & B & G & G & B & B & G & G & C & D & D & D & 0.8264 \\ 
      & B & B & B & G & G & B & B & B & D & D & C & D & 0.2500 \\ 
      & B & B & G & G & G & B & B & B & D & D & C & D & 0.2500 \\
      \hline
    \end{tabular}
 \caption{Coherent stable strategies and their normalized average payoffs
$\widetilde{W}_H$ for the case $b/c=2$ and $\eA=0.01$. The top eight
strategies (labeled Ia through to IIIb) are the so-called Leading
Eight \cite{ohtsuki:2004}. They are the ones with the highest payoffs
among all the stable strategies obtained for a given benefit-to-cost
ratio ($b/c$).
\label{tab:samemoral}}
  \end{center}
\end{table*}

On the other hand, payoff and coherence seem to be correlated. Specifically,
stable strategies with high payoff are highly coherent (incoherent for the
their mirror strategies). In Table~\ref{tab:samemoral} we list all coherent
stable strategies along with their payoffs for $b/c=2$ and $\eA=0.01$. Most of
them coincide with those found by~\cite{ohtsuki:2004}. There are some minor
differences though because we are using slightly different models. The eight
strategies with the highest payoff correspond to the so-called \textit{Leading
Eight} \cite{ohtsuki:2004}. These strategies are present in all cases shown
in Fig.~\ref{fig:samemoral}. All stable strategies have some common features:
(i) not helping good individuals is always considered bad, (ii) good
individuals never help bad ones, (iii) good individuals always help good
individuals---except when errors occur---and that is judged as good. (There
are two strategies for which the last feature is quite the opposite,
but they receive rather low payoffs.)

Notice that the absence of errors in the moral assessments renders all
defective strategies (strategies that always defect) vulnerable to invasions.
Ohtsuki and Iwasa \cite{ohtsuki:2004} found that all defective strategies were
stable; the reason is that although these strategies never reward, errors in
judgments provide them some payoff. This does not happen in the present model.
Thus defective strategies are no longer stable.

Once identified the strategies that cannot be invaded by others with the same
moral assessments, we study which of them are actually stable against the
invasion by any other strategy. We have found that all those strategies remain
stable even if strategies with different moral assessments try to invade them.
Besides, we have also checked that strategies that can be invaded by other
strategies with the same moral assessment can be invaded by some strategies
with different moral assessment as well.
%Therefore successful invasions can only be achieved by individuals sharing the
%same moral assessments, but using different action rules.

\subsection{Robustness against initial misjudgments}

We have checked sensitivity of these results with respect to a different choice
of the initial conditions to solve Eqs.~\eqref{eq:evol_x1} and
\eqref{eq:evol_x2}. In Sec.~\ref{sec:stability} we made the assumption that,
before a mutation occurs, all individuals share the same opinion about
everybody because the population is homogeneous. Initial misjudgments can lead
a fraction of the population to disagree from the general opinion. This choice
for initial conditions may be modeled as
\begin{equation}
\begin{split}
&x_i^{GG}(0)=(1-\epsilon_r^B) x_{iH}, \\
&x_i^{GB}(0)=\epsilon_r^B x_{iH}, \\
&x_i^{BB}(0)=(1-\epsilon_r^G) (1-x_{iH}), \\
&x_i^{BG}(0)=\epsilon_r^G (1-x_{iH}),
\end{split}
\label{eq:ICrumor}
\end{equation}
where $\epsilon_r^B$ ($\epsilon_r^G$) is the fraction of individuals that are
misjudged as bad (good) by the mutants. Note that if
$\epsilon_r^B=\epsilon_r^G=0$ the whole population agrees in its judgments and
we recover the former initial conditions.

Depending on the (small) values of $\epsilon_r^B$ and $\epsilon_r^G$, we have
checked that the initial conditions \eqref{eq:ICrumor} may lead to three
different scenarios. In the first one $x_i^{GG}=x_{iH}$, $x_i^{BB}=1-x_{iH}$
and $x_i^{GB}=x_i^{BG}=0$, so that misjudgments fade away and we recover a
homogeneous population. In the second scenario initial misjudgments remain or
even grow ($x_i^{GG}$ and $x_i^{BB}$ decrease and $x_i^{GB}$ and $x_i^{BG}$
increase), but the payoff obtained by the mutants is lower than that obtained
by the residents. Consequently the mutants are expelled and a homogeneous
population is restored. In the third scenario initial misjudgments also remain
and the mutants obtain higher payoffs than the residents, so that misjudgments
eventually spread. We have found that around $850$ strategies lie in this last
case (considering differences between mutant's and resident's payoffs higher
than $10^{-6}$) when $b/c=2$ and $\eA=0.01$. Fortunately none of these
strategies belong to the group of the stable ones, so this misjudgment
spreading does not affect the evolutionary fate of the population.

\subsection{Stability in the presence of cheating}

Consider now the situation in which actions are not always witnessed; instead,
there is a chance that they pass unnoticed by the rest of the population. In
this situation individuals may have the temptation to cheat by defecting
regardless of their action rules. The appearance of this kind of mutation
introduces a new set of strategies, parameterized by the cheating probability
$p_{\text{ch}}$, which might render unstable strategies that would otherwise
resist invasions. The stability will of course be a function of the probability
that the action is witnessed, $p_{\text{dis}}$.

To address this issue let us consider that residents decide to cheat with
a probability $p_{\text{ch},1}$ and mutants do so with a probability
$p_{\text{ch},2}$, in the hope that they are not discovered. However their
cheating will actually be discovered with a probability $p_{\text{dis}}$.
Assuming the same moral assessments and action rules for all individuals,
$x_1^{G*}=x_1^{*G}=x^{\text{ch}}_{1H}$ and $x_2^{G*}=x_2^{*G}=x^{\text{ch}}_{2}$,
where the fractions $x^{\text{ch}}_{1H}$ and $x^{\text{ch}}_{2}$ are calculated
as above from Eqs.~\eqref{eq:2grade} and \eqref{eq:x2G}, but incorporating the
probability of being discovered if they cheat. Likewise $P_{i,\alpha\beta}$
in Eq.~\eqref{eq:P-homo} has to be replaced by
\begin{equation}
P^{\text{ch}}_{i,\alpha\beta} = (1-p_{\text{dis}}p_{\text{ch},i})P_{i,\alpha\beta}
+p_{\text{dis}}p_{\text{ch},i}m_{i\alpha\beta}(D),
\label{eq:Pcheat}
\end{equation}
which expresses the fact that nothing changes if player $i$ either does not
cheat or she does without being discovered [probability $1-p_{\text{ch},i}+
p_{\text{ch},i}(1-p_{\text{dis}})=1-p_{\text{dis}}p_{\text{ch},i}$]; otherwise
[probability $p_{\text{dis}}p_{\text{ch},i}$] she is judged good or bad
according to $m_{i\alpha\beta}(D)$.

Finally, the probabilities of cooperation [c.f.~Eq.~\eqref{eq_theta}] are
modified as
\begin{equation}
\begin{split}
\theta^{\text{ch}}_{1,1} &=(1-p_{\text{ch},1})(1-\eA)
\sum_{\alpha\beta} \chi_\alpha(x^{\text{ch}}_{1H})
\chi_\beta(x^{\text{ch}}_{1H})\, a_{1\alpha\beta}, \\
\theta^{\text{ch}}_{1,2} &=(1-p_{\text{ch},1})(1-\eA)
\sum_{\alpha\beta} \chi_\alpha(x^{\text{ch}}_{1H})
\chi_\beta(x^{\text{ch}}_{2})\, a_{1\alpha\beta}, \\
\theta^{\text{ch}}_{2,1} &=(1-p_{\text{ch},2})(1-\eA)
\sum_{\alpha\beta} \chi_\alpha(x^{\text{ch}}_{2})
\chi_\beta(x^{\text{ch}}_{1H})\, a_{2\alpha\beta}, \\
\theta^{\text{ch}}_{2,2} &=(1-p_{\text{ch},2})(1-\eA)
\sum_{\alpha\beta} \chi_\alpha(x^{\text{ch}}_{2})
\chi_\beta(x^{\text{ch}}_{2})\, a_{2\alpha\beta}.
\end{split}
\label{eq_theta_cheat}
\end{equation}

\begin{figure*}[!ht]
\begin{center}
\includegraphics[width=7in]{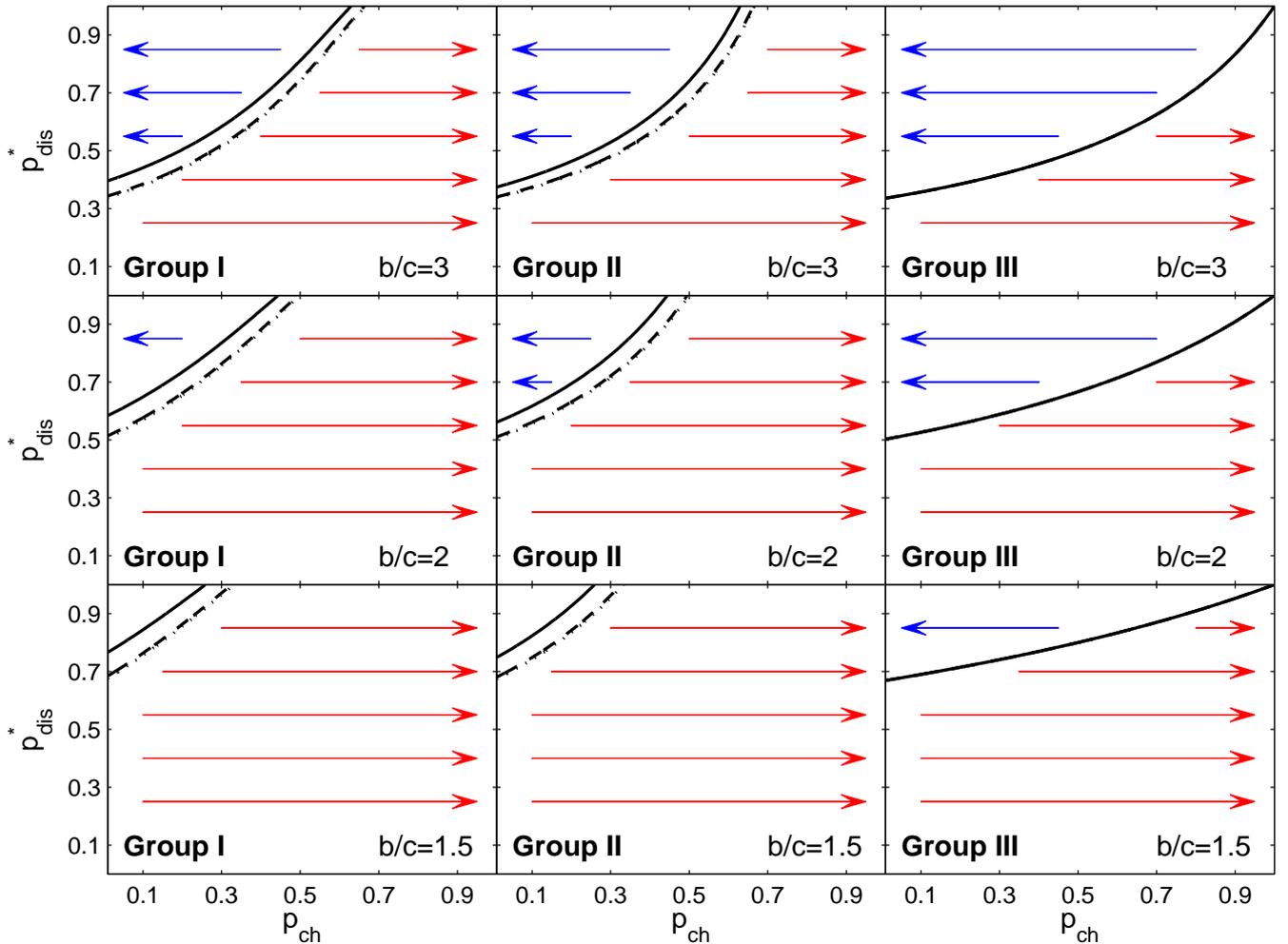}
\caption{(Color online) Limit curves of $p^*_{\text{dis}}$ as a function of
$p_{\text{ch}}$ that divide the regions where the cheating can be increased
(above the curves) and decreased (below the curves) through the invasion of
mutants with different $p_{ch}$. Different types of lines represent different
values of $\eA$: $0.1$ (continuous), $0.01$ (dashed) and $0.001$ (doted).
Different panels show results for different groups of leading eight strategies
and $b/c$. 
\label{fig:cheat}}
\end{center}
\end{figure*}

We have studied the stability of the leading eight strategies against the
invasion of cheaters. We divided the leading eight strategies in Groups I, II
and III as a function of its different behavior (as it was done in
\cite{ohtsuki:2004}). Figure~\ref{fig:cheat} represents the limiting
$p^{*}_{\text{dis}}$ below (above) which mutants who cheat with a higher
(lower) probability than residents can invade. In Appendix B we calculate
analytically the shape of this curve in the limit $\eA\to 0$.
Figure~\ref{fig:cheat} shows that below the curve
$p^{*}_{\text{dis}}(p_{\text{ch}})$ cheating increases without bound through
subsequents invasions until the whole population is dominated by defectors.
In other words, if cheating occurs and the probability of being discovered is
not high enough, none of the leading eight strategies survives. In particular,
if $p_{\text{dis}}<c/b$ full defection is the unavoidable fate of the
population. Thus, if only small mutations are allowed in a honest population,
we find the Hamilton-like rule $bp_{\text{dis}}>c$ for the survival of
cooperation \cite{hamilton:1964a}.

Increasing $\eA$ makes it even easier for cheaters to invade, with the exception
of the strategies belonging to Group III, which seem to be insensitive to the
effect of errors in action.

\section{Discussion}
\label{sec:discussion}

We have carried out a systematic study of the stability of all possible
third-order indirect reciprocity strategies. We extended the work of
Ohtsuki and Iwasa \cite{ohtsuki:2004} confronting all the strategies against
the others regardless of whether they have the same moral assessments or not.
%In order do that, we used a variant of the model studied in
%Ref.~\cite{ohtsuki:2004}.
The main difference with their model is that in ours individuals
directly witness all actions. Allowing individuals in the same population to
have different moral assessments and action rules makes indirect observation
models computationally unfeasible (we must store everybody's opinion of
everybody else at every time step). For the same reason, errors in judgments
cannot be accounted for in direct observation models. Thus we only consider
errors in performing the actions. The only exception to this assumption is the
need to introduce errors in judgement to calculate, in some special cases,
the stationary fractions of good and bad individuals in homogeneous populations.
But this is just a technical issue that allows us to resolve a degeneracy of
solutions, and there is no inconsistency because the results do not depend on
the value of this error.

The strategies which are stable against invasions by other strategies sharing
the same moral assessment turn out to be also stable against invasions by any
other strategy. This means that if a strategy can resist the invasion of all
the other strategies that share its same moral assessment, it can resits any
invasion whatsoever.

We have checked that the higher the benefit-to-cost ratio and the lower the
action errors the higher the number of stable strategies obtained. One
possible interpretation of the errors in action is lack of resources. Under
this interpretation our results show that scarcity of resources favors
invasions. On the other hand, we have checked that populations whose members
receive more benefit for a given cost are more resistant to invasions.

As pointed out in Ref.~\cite{ohtsuki:2004}, there is a symmetry between the
moral assessments of the strategies. Good and bad are just labels with no
proper meaning---in contrast to actions, that have a direct influence in the
payoffs. In order to break that symmetry and provide a meaning to those
labels we have introduced the concept of coherence. Coherence links moral
assessments with action rules. We have shown that stable strategies appear in
pairs due to the above mentioned symmetry, but coherence values are
complementary. This allows us to choose only one of the strategies (the most
coherent) within each pair for later analysis and interpretation.

The stable strategies we obtain include the Leading Eight found by Ohtsuki and
Iwasa \cite{ohtsuki:2004}. These are also the most efficient ones (those with
highest payoffs). Both the Leading Eight as well as the remaining stable
strategies that we have obtained share some features, and except for the two
least efficient strategies (with $\widetilde{W}_H=0.25$), all of them obtain
high average payoffs ($\widetilde{W}_H>0.8$).
%All of them maintain cooperation
%($a_{GG}=C$ and $m_{GG}(C)=G$) and identify defectors ($m_{GG}(D)=m_{BG}=B$).
%They also
They identify defectors ($m_{GG}(D)=m_{BG}(D)=B$) and, except the two least
efficient strategies, maintain cooperation ($a_{GG}=C$ and $m_{GG}(C)=G$). All
of them punish defectors ($a_{GB}=D$), although three of the stable
strategies (with $\widetilde{W}_H\sim 0.9$) do not judge this as a good
behavior. Finally the most efficient stable strategies ($\widetilde{W}_H>0.9$)
forgive bad individuals who help good players ($m_{BG}(C)=G$ and $a_{BG}=C$).
The more of these features the strategies follow the higher their payoff. For
instance, the three strategies with $\widetilde{W}_H\sim 0.9$ turn good
punishers into bad individuals and they can only restore their reputation by
helping good individuals. And in the case of strategies with
$\widetilde{W}_H<0.9$, bad individuals cannot increase their reputation by
helping good players, but only by interacting with other bad individuals.

We have also found that all these strategies may become unstable if cheaters
arise. If the probability of witnessing a cheat is not high enough, cheaters
can take over an honest population. Upon increasing the cheating probability
$p_{\text{dis}}>c/b$ the population eventually turns into pure defectors.
Interestingly, the condition for a population to resist this effect is of the
Hamilton type, namely, $bp_{\text{dis}}>c$, where $b$ is the benefit and $c$
the cost. Errors in action make this condition even more restrictive for the
stability of a honest population.

Cheating is always a danger for cooperation based on indirect reciprocity.
Even in societies where this mechanism is of utmost importance cheating always
threats honest behavior. For instance, the (now extinct) Patagonian tribes
of the Y\'amana are among the reported societies more strongly based on
indirect reciprocity \cite{orquera:1999}. Sharing food even with nonrelatives
appeared to be the default behavior. Not sticking to it brought a bad reputation
and severe social pubishment (e.g., not participating in further food sharing).
Yet, cheating among the Y\'amana was reported to occur when chances were low
to be discovered (for instance, because the prey obtained was easy to hide;
see Ref.~\cite{orquera:1999}, p.~197).

One of the problems that emerges from considering different moral assessments
is the possibility that the fractions of good and bad individuals may depend on
the initial setup. We sort out this issue by choosing realistic initial
conditions for the differential equations describing the evolution of these
fractions. Essentially, we assume that mutations do not change the previous
judgments that individuals had on each other. This notwithstanding, we have
analyzed other initial conditions in which not all individuals have the same
opinion. A typical setup where this might happen is when a rumor is spread
over a fraction of the population. We have checked that, although misjudgement
can survive or even spread over a larger fraction of the population, it
eventually disappears because mutants with a wrong judgement get less payoff
than residents who use one of the stable strategies.

Admittedly, in order to carry out such a systematic analysis as we have
performed here, we have had to sacrifice some realism in the model. On the one
hand, we have considered that reputation can only have two states: good and
bad. This binary reputation have been used in several preceding studies
\cite{ohtsuki:2004, uchida:2010} and implies that only the actions that happen
in the last round are taken into account to assign reputation. However,
Tanabe et al.~\cite{tanabe:2012} have studied a model with trinary reputations
and showed that some strategies (like the so-called \textit{image scoring)} can
be stable in a trinary-reputation model but not in a binary-reputation one. On
the other hand, we have considered that every player has complete information
of every single interaction in the population (except when we introduced
cheating). This is too strong an assumption and some studies discuss the effect
of a limited access to the information (see~\cite{nakamura:2011} and the
references therein).

\section*{Acknowledgments}

We are indebted to Prof.~Karl Sigmund for his hospitality in hosting LAMV
and his valuable insights. We also thank Ivan Briz for sharing his
archaeological knowledge with us. This work has been supported by Ministerio de
Ciencia e Innovaci\'on (Spain) through grants MOSAICO and PRODIEVO, by European
Research Area Complexity-Net through grant RESINEE, and by Comunidad de Madrid
(Spain) through grant MODELICO-CM. LAMV was supported by a postdoctoral
fellowship from Alianza 4 Universidades.

\appendix

\section{}
\label{sec:appendixA}

The two sets of Eqs.~\eqref{eq:evol_x1} and \eqref{eq:evol_x2} can be
simplified in the steady state $dx/dt=0$. Thus, summing over the reputation
$\Lambda_2$ in Eqs.~\eqref{eq:evol_x1} we obtain
\begin{equation}
x_{1}^{G*}=\sum_{\Lambda_2} x_1^{G\Lambda_2}=x_{1H}.
\end{equation}
Therefore we can reduce Eqs.~\eqref{eq:evol_x1} to just two equations in two
unknowns (e.g., $x_{1}^{GG}$ and $x_{1}^{BB}$) by setting
\begin{equation}
\begin{split}
&x_{1}^{GB}=x_{1}^{G*}-x_{1}^{GG}, \\
&x_{1}^{BG}=1-x_{1}^{G*}-x_{1}^{BB}.
\end{split}
\label{eq_x1BG}
\end{equation}
The two remaining equations from \eqref{eq:evol_x1} have to be solved
numerically using the initial conditions discussed in Sec.~\ref{sec:stability}.

On the other hand, the set of Eqs.~\eqref{eq:evol_x2} is decoupled from the set
\eqref{eq:evol_x1}, and so they can be solved analytically after solving the
latter. This is easier if $x_2^{*G}$ is calculated first,
\begin{equation}
\begin{split}
x_2^{*G}=& \left[ x_{1}^{*G}P_{2,01} + (1-x_{1}^{*G})P_{2,00}\right]  \\
&\times\left[ 1 + x_{1}^{*G}(P_{2,01}-P_{2,11}) \right. \\
&\left. + (1-x_{1}^{*G})(P_{2,00}-P_{2,10}) \right]^{-1}.
\end{split}
\label{eq:x2G}
\end{equation}
Hence Eq.~\eqref{eq:evol_x2} reduces to a linear system of two equations
in the two unknowns $x_2^{GG}$, $x_2^{BB}$.

There are scenarios where the solution of $x_2^{\Lambda_1\Lambda_2}$ turns out
to be degenerated. In these situations the set of Eqs.~\eqref{eq:evol_x2} need
to be integrated along with the set of Eqs.~\eqref{eq:evol_x1}.

\section{}
\label{sec:appendixB}

Consider a resident population whose individuals play one of the leading eight
strategies with probability $1-p_{\text{\text{ch}},1}$ but defect otherwise.
Consider mutants who do the same, but with a probability $1-p_{\text{ch},2}$.
For simplicity let us assume the limiting case $\eA\to 0$. Applying adaptive
dynamics \cite{hofbauer:1998}, the curve separating the regions where the
mutant can or cannot invade the population is given by
\begin{equation}
 \left. \frac{dW(p_{\text{\text{ch}},2},p_{\text{ch},1})}{dp_{\text{ch},2}}
\right|_{p_{\text{ch},2}=p_{\text{ch},1}}=0,
\label{eq:adapdyn}
\end{equation}
where the payoff $W(p_{\text{\text{ch}},2},p_{\text{ch},1})$ is equivalent to
$W(2|1)$. According to Eq.~\eqref{eq_W22},
\begin{equation}
\frac{dW(p_{\text{\text{ch}},2},p_{\text{ch},1})}{dp_{\text{ch},2}} =
b\frac{d\theta^{ch}_{1,2}}{dp_{\text{ch},2}} 
-c\frac{d\theta^{ch}_{2,1}}{dp_{\text{ch},2}}.
\end{equation}

To go further we need to separate the strategies of the three groups.

\subsection{Group I strategies}

Using Eqs.~\eqref{eq_theta_cheat} for the leading eight strategies, the
probabilities of cooperation $\theta^{\text{ch}}_{i,j}$ are
\begin{equation}
\begin{split}
\theta^{\text{ch}}_{1,2}=&(1-p_{\text{\text{ch}},1})
\left[ x_{2}^{\text{ch}}+(1-x_{1,H}^{\text{ch}})(1-x_{2}^{\text{ch}}) \right], \\
\theta^{\text{ch}}_{2,1}=&(1-p_{\text{\text{ch}},2})
\left[ x_{1,H}^{\text{ch}}+(1-x_{1,H}^{\text{ch}})(1-x_{2}^{\text{ch}})\right].
\end{split}
\label{eq:thetachL8I}
\end{equation}
Thus
\begin{equation}
\begin{split}
\dfrac{d\theta^{\text{ch}}_{1,2}}{dp_{\text{ch},2}}=&
(1-p_{\text{\text{ch}},1})\,x_{1,H}^{ch}\,
\dfrac{dx_{2}^{ch}}{dp_{\text{ch},2}}, \\
\dfrac{d\theta^{\text{ch}}_{2,1}}{dp_{\text{ch},2}}=&
(1-x_{1,H}^{\text{ch}})\left[ x_{2}^{ch}-(1-p_{\text{ch},2})\,
\dfrac{dx_{2}^{ch}}{dp_{\text{ch},2}} \right] -1.
\end{split}
\label{eq:DthetachL8I}
\end{equation}

The fractions $x_{1,H}^{\text{ch}}$ and $x_{2}^{ch}$ are obtained from
Eqs.~\eqref{eq:2grade} and \eqref{eq:x2G}. To that purpose we need to
substitute
\begin{equation}
P_{i,11}^{\text{ch}}= P_{i,01}^{\text{ch}}=
 1-p_{\text{dis}}p_{\text{ch},i}, \qquad P_{i,10}^{\text{ch}}= 1
\label{eq:PchL8I}
\end{equation}
and
\begin{equation}
P_{i,00}^{\text{ch}}=1-p_{\text{dis}}p_{\text{ch},i}.
\label{eq:PchL8I00}
\end{equation}
Thus $x_{1,H}^{\text{ch}}$ is the solution of
\begin{equation}
p_{\text{dis}}p_{\text{ch},1}(x_{1,H}^{\text{ch}})^2
=(1-p_{\text{dis}}p_{\text{ch},1})(1-x_{1,H}^{\text{ch}}),
\end{equation}
and once it is obtained,
\begin{equation}
\begin{split}
&x_{2}^{\text{ch}}= \frac{1-p_{\text{dis}}p_{\text{ch},2}}
{1-(1-x_{1,H}^{\text{ch}})p_{\text{dis}}p_{\text{ch},2}}, \\
&\dfrac{dx_{2}^{\text{ch}}}{dp_{\text{ch},2}}=
-\frac{p_{\text{dis}}x_{1,H}^{\text{ch}}}{[1-(1-x_{1,H}^{\text{ch}})
p_{\text{dis}}p_{\text{ch},2}]^2}.
\end{split}
\end{equation}
Substituting into \eqref{eq:DthetachL8I} and setting $p_{\text{ch},2}
=p_{\text{ch},1}\equiv p_{\text{ch}}$ yields
\begin{equation}
\begin{split}
\dfrac{d\theta^{\text{ch}}_{1,2}}{dp_{\text{ch}}}
=& -\frac{(1-p_{\text{ch},1})p_{\text{dis}}
(x_{1,H}^{\text{ch}})^2}{[1-p_{\text{dis}}p_{\text{ch}}(1-x_{1,H}^{\text{ch}})]^2}, \\
\dfrac{d\theta^{\text{ch}}_{2,1}}{dp_{\text{ch},2}}
=& \frac{x_{1,H}^{\text{ch}}[p_{\text{dis}}(1-x_{1,H}^{\text{ch}})-1]}
{[1-p_{\text{dis}}p_{\text{ch}}(1-x_{1,H}^{\text{ch}})]^2}.
\end{split}
\end{equation}
Therefore $p_{\text{dis}}^*$ is the solution of the system
\begin{equation}
\begin{split}
& p_{\text{dis}}^*[b(1-p_{\text{ch}})x^*+c(1-x^*)]=c, \\
& p_{\text{dis}}^*p_{\text{ch}}(x^*)^2=(1-p_{\text{dis}}^*p_{\text{ch}})(1-x^*).
\end{split}
\end{equation}

\subsection{Group II strategies}

For the strategies of this group
\begin{equation}
\theta^{\text{ch}}_{1,2}= (1-p_{\text{\text{ch}},1})\,x_{2}^{\text{ch}}, \qquad
\theta^{\text{ch}}_{2,1}= (1-p_{\text{\text{ch}},2})\,x_{1,H}^{\text{ch}},
\label{eq:thetachL8II}
\end{equation}
hence their derivatives are
\begin{equation}
\dfrac{d\theta^{\text{ch}}_{1,2}}{dp_{\text{ch},2}}=
(1-p_{\text{\text{ch}},1})\,\dfrac{dx_{2}^{ch}}{dp_{\text{ch},2}}, \qquad
\dfrac{d\theta^{\text{ch}}_{2,1}}{dp_{\text{ch},2}}= -x_{1}^{\text{ch}}.
\label{eq:DthetachL8II}
\end{equation}
Probabilities $P_{i,\alpha\beta}^{\text{ch}}$ are now given by
\eqref{eq:PchL8I} as well as $P_{i,00}^{\text{ch}}=1$. Thus, after
Eqs.~\eqref{eq:2grade} and \eqref{eq:x2G},
\begin{equation}
x_{1,H}^{\text{ch}}=\frac{1}{1+p_{\text{dis}}p_{\text{\text{ch}},1}}, \qquad
x_2^{\text{ch}}=1-x_{1,H}^{\text{ch}}p_{\text{dis}}p_{\text{\text{ch}},2}.
\end{equation}
Substituting into \eqref{eq:DthetachL8II} and setting $p_{\text{ch},2}
=p_{\text{ch},1}\equiv p_{\text{ch}}$ yields
\begin{equation}
\begin{split}
\dfrac{d\theta^{\text{ch}}_{1,2}}{dp_{\text{ch},2}}
=& -\frac{(1-p_{\text{ch}})p_{\text{dis}}}{1+p_{\text{dis}}p_{\text{ch}}}, \\
\dfrac{d\theta^{\text{ch}}_{2,1}}{dp_{\text{ch,2}}}
=& -\frac{1}{1+p_{\text{dis}}p_{\text{ch}}},
\end{split}
\end{equation}
and therefore
\begin{equation}
p_{\text{dis}}^*=\dfrac{c}{b\,(1-p_{\text{ch}})}.
\end{equation}

\subsection{Group III strategies}

For the strategies of this group the probabilities of cooperation and their
derivatives are also given by Eqs.~\eqref{eq:thetachL8II} and
\eqref{eq:DthetachL8II}, and the probabilities $P_{i,\alpha\beta}^{\text{ch}}$
by \eqref{eq:PchL8I} as well as $P_{i,00}^{\text{ch}}=0$. Thus, after
Eqs.~\eqref{eq:2grade} and \eqref{eq:x2G},
\begin{equation}
x_{1,H}^{\text{ch}}=1-p_{\text{dis}}p_{\text{\text{ch}},1}, \qquad
x_2^{\text{ch}}=1-p_{\text{dis}}p_{\text{\text{ch}},2}.
\end{equation}
Substituting into \eqref{eq:DthetachL8II} and setting $p_{\text{ch},2}
=p_{\text{ch},1}\equiv p_{\text{ch}}$ yields
\begin{equation}
\begin{split}
\dfrac{d\theta^{\text{ch}}_{1,2}}{dp_{\text{ch},2}}
=& -(1-p_{\text{ch}})p_{\text{dis}}, \\
\dfrac{d\theta^{\text{ch}}_{2,1}}{dp_{\text{ch,2}}}
=& -(1-p_{\text{dis}}p_{\text{ch}}),
\end{split}
\end{equation}
and therefore
\begin{equation}
p_{\text{dis}}^*=\dfrac{c}{c\,p_{\text{ch}}+b\,(1-p_{\text{ch}})}.
\end{equation}

\bibliographystyle{apsrev4-1}
\bibliography{evol-coop}

\end{document}